\documentstyle[11pt,epsfig]{article}

\begin{document}

\vspace{0.2cm}

\begin{center}
{\large \bf Neutrino Mixing Angles from Texture Zeros of the Lepton
Mass Matrices}
\end{center}

\vspace{0.2cm}
\begin{center}
{\bf Harald Fritzsch} \footnote{E-mail: fritzsch@mppmu.mpg.de} \\
{\sl Department f\"{u}r Physik, Universit\"{a}t M\"{u}nchen,
Theresienstra{\ss}e 37, 80333 M\"{u}nchen, Germany} \\

~\\

{\bf Shun Zhou} \footnote{E-mail: shunzhou@kth.se} \\
{\sl Department of Theoretical Physics, School of Engineering
Sciences, KTH Royal Institute of Technology, AlbaNova University
Center, Roslagstullsbacken 21, 106 91 Stockholm, Sweden}
\end{center}

\vspace{2.0cm}

\begin{abstract}
Taking into account the latest neutrino oscillation data, we study
texture zeros of the lepton mass matrices. Assuming the Dirac
neutrino mass matrix $M^{}_{\rm D}$, the charged-lepton mass matrix
$M^{}_l$ and the mass matrix of heavy right-handed Majorana
neutrinos $M^{}_{\rm R}$ to have three texture zeros, we show that
the observed neutrino mixing angles can naturally be obtained. The
phenomenological implications for the neutrino mass spectrum, the
CP-violating phases, the tritium beta decay and the neutrinoless
double-beta decay are explored.
\end{abstract}

\begin{center}
{\small PACS number: 14.60.Pq}
\end{center}

\newpage

\section{Introduction}

The recent Daya Bay \cite{Daya} and RENO \cite{Reno} reactor
neutrino experiments reveal that the smallest neutrino mixing angle
is relatively large (i.e., $\theta^{}_{13} \approx 9^\circ$). The
solar and atmospheric neutrino oscillation experiments indicate that
the other two mixing angles are large (i.e., $\theta^{}_{12} \approx
34^\circ$ and $\theta^{}_{23} \approx 40^\circ$). This has been
confirmed by the long-baseline accelerator and reactor neutrino
oscillation experiments. Two independent neutrino mass-squared
differences have been determined: $\delta m^2 \equiv m^2_2 - m^2_1
\approx 7.5\times 10^{-5}~{\rm eV}^2$ and $\Delta m^2 \equiv m^2_3 -
(m^2_1 + m^2_2)/2 \approx \pm 2.4 \times 10^{-3}~{\rm eV}^2$, where
$m^{}_i$ (for $i=1, 2, 3$) are the neutrino masses. The next
important step in neutrino physics will be the determination of the
sign of $\Delta m^2$ and the measurement of the leptonic
CP-violating phase.

It has been shown that the texture zeros in the quark mass matrices
\cite{F}
\begin{equation}
M^{}_q = \left(\matrix{ 0 & \times & 0 \cr \times & 0 & \times \cr 0
& \times & \times} \right) \; ,
%     (1)
\end{equation}
give successful relations between the quark mixing angles and the
quark mass ratios. The texture zeros were used for the lepton mass
matrices to explain the observed bi-large neutrino mixing pattern,
given a weakly-hierarchical neutrino mass spectrum (e.g., $m^{}_1 :
m^{}_2 : m^{}_3 \approx 1 : 3 : 10$) \cite{Xing02,XZ04}.

We shall reexamine the texture zeros and confront them with the
recent neutrino oscillation data. It turns out the original model,
where both the charged-lepton and effective neutrino mass matrices
assume the same texture zeros, has been already ruled out, since the
observed relatively large mixing angle $\theta^{}_{13}$ requires a
large ratio of two neutrino mass-squared differences, which is not
allowed by the current data. We consider the canonical seesaw model
and use the texture zeros for the Dirac neutrino mass matrix
$M^{}_{\rm D}$, the charged-lepton mass matrix $M^{}_l$, and the
mass matrix of the heavy right-handed Majorana neutrinos $M^{}_{\rm
R}$. Then the effective neutrino mass matrix is given by the
well-known seesaw formula $M^{}_\nu = M^{}_{\rm D} M^{-1}_{\rm R}
M^T_{\rm D}$, and both a nonmaximal $\theta^{}_{23}$ and an
unsuppressed $\theta^{}_{13}$ can be naturally accommodated.

This paper is organized as follows. In Sec.~2 we apply the texture
zeros to the effective neutrino mass matrix as well as the charged
lepton mass matrix. We explain why this scenario is now disfavored
by current neutrino oscillation data. Sec.~3 is devoted to a
canonical seesaw model with texture zeros of the lepton mass
matrices. Two models of the effective neutrino mass matrix have been
discussed in detail. The phenomenological implications are explored,
including the neutrino mixing angles, the neutrino mass spectrum,
the leptonic CP violation, the tritium beta decay and the
neutrinoless double-beta decays. We summarize the conclusions in
Sec. 4.

\section{Effective Neutrino Mass Matrix}

The lepton mass spectra and mixing parameters are determined by the
lepton mass terms
\begin{equation}
-{\cal L}^{}_{\rm m} = \overline{\left(\matrix{e^{}_{\rm L} &
\mu^{}_{\rm L} & \tau^{}_{\rm L}}\right)} M^{}_l
\left(\matrix{e^{}_{\rm R} \cr \mu^{}_{\rm R} \cr \tau^{}_{\rm
R}}\right) + \frac{1}{2} \overline{\left(\matrix{\nu^{}_{e \rm L} &
\nu^{}_{\mu \rm L} & \nu^{}_{\tau \rm L}}\right)} M^{}_\nu
\left(\matrix{\nu^{\rm C}_{e \rm L} \cr \nu^{\rm C}_{\mu \rm L} \cr
\nu^{\rm C}_{\tau \rm L}}\right) + {\rm h.c.} \; ,
%     (2)
\end{equation}
with $\nu^{\rm C}_{\alpha \rm L} \equiv C \overline{\nu^{}_{\alpha
\rm L}}^{\rm T}$ (for $\alpha = e, \mu, \tau$), where $M^{}_l$ and
$M^{}_\nu$ stand for the charged-lepton mass matrix and the
effective neutrino mass matrix. Since $M^{}_\nu$ is a symmetric
complex matrix, we take $M^{}_l$ to be symmetric as well.
Furthermore we assume that both $M^{}_\nu$ and $M^{}_l$ are of the
following form:
\begin{equation}
M^{}_f = \left(\matrix{ 0 & {\cal C}^{}_f & 0 \cr {\cal C}^{}_f & 0
& {\cal B}^{}_f \cr 0 & {\cal B}^{}_f & {\cal A}^{}_f}\right) \; ,
%     (3)
\end{equation}
where ${\cal C}^{}_f$ and ${\cal B}^{}_f$ are in general complex,
while ${\cal A}^{}_f$ can be made real by removing the overall phase
of $M^{}_f$. Since both $M^{}_l$ and $M^{}_\nu$ are symmetric, they
can be diagonalized through a unitary transformation $U^\dagger_f
M^{}_f U^*_f = {\rm Diag}\{\lambda^f_1, \lambda^f_2, \lambda^f_3\}$,
where $\lambda^f_i$ (for $i = 1, 2, 3$) denote the lepton mass
eigenvalues. After the diagonalization of $M^{}_l$ and $M^{}_\nu$
the lepton mixing matrix is given by $V = U^\dagger_l U^{}_\nu$. The
diagonalization can be done as follows. The mass matrix $M^{}_f$ in
Eq.~(3) can be decomposed into $M^{}_f \equiv P^{}_f
\overline{M}^{}_f P^T_f$ with
\begin{equation}
~~~ \overline{M}^{}_f = \left(\matrix{ 0 & C^{}_f & 0 \cr C^{}_f & 0
& B^{}_f \cr 0 & B^{}_f & A^{}_f}\right) \;
%     (4)
\end{equation}
and $P^{}_f = {\rm Diag}\{e^{i(\varphi^{}_f - \phi^{}_f)},
e^{i\phi^{}_f}, 1\}$, where $A^{}_f = {\cal A}^{}_f$, $B^{}_f =
|{\cal B}^{}_f|$, $C^{}_f = |{\cal C}^{}_f|$, $\phi^{}_f =
\arg[{\cal B}^{}_f]$ and $\varphi^{}_f = \arg[{\cal C}^{}_f]$. Then
the real and symmetric matrix $\overline{M}^{}_f$ is diagonalized:
\begin{equation}
(O^{}_f Q)^T \overline{M}^{}_f (O^{}_f Q) = \left(\matrix{
\lambda^f_1 & 0 & 0 \cr 0 & \lambda^f_2 & 0 \cr 0 & 0 &
\lambda^f_3}\right) \; ,
%     (5)
\end{equation}
where $Q = {\rm Diag}\{1, i, 1\}$ has been introduced to cancel the
minus sign of ${\rm Det}[\overline{M}^{}_f] = - A^{}_f C^2_f$, and
$O^{}_f$ is a real orthogonal matrix. Therefore the nonzero matrix
elements of $\overline{M}^{}_f$ can be expressed in terms of the
eigenvalues $\lambda^f_i$ (for $i=1, 2, 3$):

\begin{eqnarray}
A^{}_f &=& \lambda^f_1 - \lambda^f_2 +
\lambda^f_3 \; , \nonumber \\
B^{}_f &=& \left[\frac{(\lambda^f_1 - \lambda^f_2)(\lambda^f_2 -
\lambda^f_3)(\lambda^f_1 + \lambda^f_3)}{\lambda^f_1 - \lambda^f_2 +
\lambda^f_3}\right]^{1/2} \; , \nonumber \\
C^{}_f &=& \left(\frac{\lambda^f_1 \lambda^f_2
\lambda^f_3}{\lambda^f_1 - \lambda^f_2 + \lambda^f_3}\right)^{1/2}
\; .
%     (6)
\end{eqnarray}
The matrix elements of the orthogonal matrix $O^{}_f$ are:

\begin{eqnarray}
O^f_{11} & = & + \left [ \frac{x^{}_f - z^{}_f} {\left (1+x^{}_{\rm
f} \right ) \left (1-z^{}_f \right ) \left (x^{}_f - z^{}_f + x^{}_f
z^{}_f \right )} \right ]^{1/2} \; ,
\nonumber \\
O^f_{12} & = & - \left [ \frac{x^3_f \left (1+z^{}_f \right )}
{\left (1+x^{}_f \right ) \left (x^{}_f + z^{}_f \right ) \left
(x^{}_f - z^{}_f + x^{}_f z^{}_f \right )} \right ]^{1/2} \; ,
\nonumber \\
O^f_{13} & = & + \left [ \frac{z^3_f \left ( 1 - x^{}_f \right )}
{\left (1-z^{}_f \right ) \left (x^{}_f + z^{}_f \right ) \left (
x^{}_f - z^{}_f + x^{}_f z^{}_f \right )} \right ]^{1/2} \; ,
\nonumber \\
O^f_{21} & = & + \left [ \frac{x^{}_f - z^{}_f} {\left (1+x^{}_{\rm
f} \right ) \left (1-z^{}_f \right )} \right ]^{1/2} \; ,
\nonumber \\
O^f_{22} & = & + \left [ \frac{ x^{}_f \left (1+z^{}_f \right
)}{\left (1+x^{}_f \right ) \left (x^{}_f + z^{}_f \right )} \right
]^{1/2} \; ,
\nonumber \\
O^f_{23} & = & + \left [ \frac{z^{}_f \left (1-x^{}_f \right )}
{\left (1-z^{}_f \right ) \left (x^{}_f + z^{}_f \right )} \right
]^{1/2} \; ,
\nonumber \\
O^f_{31} & = & - \left [ \frac{x^{}_f z^{}_f \left (1-x^{}_f \right
) \left (1+z^{}_f \right )}{\left (1+x^{}_f \right ) \left (1-z^{}_f
\right ) \left (x^{}_f - z^{}_f + x^{}_f z^{}_f \right )} \right
]^{1/2} \; ,
\nonumber \\
O^f_{32} & = & - \left [ \frac{z^{}_f \left (1 -x^{}_f \right )
\left (x^{}_f - z^{}_f \right )}{\left (1+x^{}_f \right ) \left
(x^{}_f + z^{}_f \right ) \left (x^{}_f - z^{}_f + x^{}_f z^{}_{\rm
f} \right )} \right ]^{1/2} \; ,
\nonumber \\
O^f_{33} & = & + \left [ \frac{x^{}_f \left (1 +z^{}_f \right )
\left (x^{}_f - z^{}_f \right )}{\left (1-z^{}_f \right ) \left
(x^{}_f + z^{}_f \right ) \left (x^{}_f - z^{}_f + x^{}_f z^{}_{\rm
f} \right )} \right ]^{1/2} \; ,
%       (7)
\end{eqnarray}
where we have defined $x^{}_f \equiv \lambda^f_1/\lambda^f_2$ and
$z^{}_f \equiv \lambda^f_1/\lambda^f_3$. For the charged leptons we
have: $x^{}_l = m^{}_e/m^{}_\mu \approx 4.84\times 10^{-3}$ and
$z^{}_l = m^{}_e/m^{}_\tau \approx 2.87 \times 10^{-4}$. For the
neutrinos we have: $x^{}_\nu = m^{}_1/m^{}_2$ and $z^{}_\nu =
m^{}_1/m^{}_3$. From Eq.~(7) one finds that only the normal neutrino
mass hierarchy with $z^{}_\nu < x^{}_\nu < 1$ is allowed. Therefore
$M^{}_l$ is diagonalized through $U^\dagger_l M^{}_l U^*_l = {\rm
Diag}\{m^{}_e, m^{}_\mu, m^{}_\tau\}$ with $U^{}_l = P^{}_l O^{}_l
Q^*$, while $M^{}_\nu$ through $U^\dagger_\nu M^{}_\nu U^*_\nu =
{\rm Diag}\{m^{}_1, m^{}_2, m^{}_3\}$ with $U^{}_\nu = P^{}_\nu
O^{}_\nu Q^*$. Except for the phases $\varphi^{}_{l, \nu}$ and
$\phi^{}_{l, \nu}$ the unitary matrices $U^{}_l$ and $U^{}_\nu$ are
determined by the mass ratios of charged leptons and neutrinos. For
the lepton mixing matrix $V = U^\dagger_l U^{}_\nu$ the absolute
values of the matrix elements can be explicitly written as
\begin{equation}
|V^{}_{p q}| = |O^l_{1p} O^\nu_{1q} e^{i\alpha} + O^l_{2p}
O^\nu_{2q} e^{i\beta} + O^l_{3p} O^\nu_{3q}| \; ,
%     (8)
\end{equation}
where $p$ and $q$ run over $1, 2, 3$. We defined $\beta \equiv
\phi^{}_\nu - \phi^{}_l$ and $\alpha \equiv (\varphi^{}_\nu -
\varphi^{}_l) - \beta$. The mixing matrix $V$ is entirely determined
by the charged-lepton mass ratios $(x^{}_l, z^{}_l)$, the neutrino
mass ratios $(x^{}_\nu, z^{}_\nu)$ and two phases $(\alpha, \beta)$.
The current experimental data on neutrino oscillations will place
restrictive constraints on these four unknown parameters $(x^{}_\nu,
z^{}_\nu)$ and $(\alpha, \beta)$.

In the standard parametrization the lepton mixing matrix $V$ is
\begin{equation}
V = \left(\matrix{c^{}_{12} c^{}_{13} & s^{}_{12} c^{}_{13} &
s^{}_{13} \cr -c^{}_{12} s^{}_{23} s^{}_{13} - s^{}_{12} c^{}_{23}
e^{-i\delta} & -s^{}_{12} s^{}_{23} s^{}_{13} + c^{}_{12} c^{}_{23}
e^{-i\delta} & s^{}_{23} c^{}_{13} \cr -c^{}_{12} c^{}_{23}
s^{}_{13} + s^{}_{12} s^{}_{23} e^{-i\delta} & -s^{}_{12} c^{}_{23}
s^{}_{13} - c^{}_{12} s^{}_{23} e^{-i\delta} & c^{}_{23}
c^{}_{13}}\right) \left(\matrix{e^{i\rho} & 0 & 0 \cr 0 &
e^{i\sigma} & 0 \cr 0 & 0 & 1}\right) \; .
%     (9)
\end{equation}
Here $s^{}_{ij} \equiv \sin \theta^{}_{ij}$ and $c^{}_{ij} \equiv
\cos \theta^{}_{ij}$, $\delta$ is the Dirac-type CP-violating phase
and $(\rho, \sigma)$ are the Majorana-type CP-violating phases. For
the normal neutrino mass hierarchy the latest global-fit analysis
yields at the $3\sigma$ level \cite{Fogli}:
\begin{eqnarray}
0.259 \leq &\sin^2 \theta^{}_{12}& \leq 0.359 \; ,
\nonumber \\
0.331 \leq &\sin^2 \theta^{}_{23}& \leq 0.637 \; ,
\nonumber \\
0.017 \leq &\sin^2 \theta^{}_{13}& \leq 0.031 \;
%     (10)
\end{eqnarray}
and
\begin{eqnarray}
6.99\times 10^{-5}~{\rm eV}^2 \leq &\delta m^2& \leq 8.18 \times
10^{-5}~{\rm eV}^2 \; ,
\nonumber \\
2.19\times 10^{-3}~{\rm eV}^2 \leq & \hspace{-0.25cm}  \Delta m^2
\hspace{-0.25cm} & \leq 2.62 \times 10^{-3}~{\rm eV}^2 \; .
%     (11)
\end{eqnarray}
Here the neutrino mass-squared differences are defined as $\delta
m^2 \equiv m^2_2 - m^2_1$ and $\Delta m^2 \equiv m^2_3 - (m^2_1 +
m^2_2)/2$. The best-fit values, $1\sigma$- and $2\sigma$-ranges of
neutrino mixing parameters can be found in Table~1. Furthermore we
have
\begin{equation}
R^{}_\nu \equiv \frac{\delta m^2}{\Delta m^2} =
\frac{z^2_\nu}{x^2_\nu} \cdot \frac{1 - x^2_\nu}{1 - z^2_\nu (1 +
x^{-2}_\nu)/2} \; ,
%     (12)
\end{equation}
and
\begin{eqnarray}
~~~~~ \sin^2 \theta^{}_{12} = \frac{|V^{}_{e2}|^2}{1 -
|V^{}_{e3}|^2} \; , ~~~~ \sin^2 \theta^{}_{23} = \frac{|V^{}_{\mu
2}|^2}{1 - |V^{}_{e3}|^2} \; , ~~~~ \sin^2 \theta^{}_{13} =
|V^{}_{e3}|^2 \; ,
%     (13)
\end{eqnarray}
where $V^{}_{\alpha i}$ should be identified as $V^{}_{j i}$ with
$\alpha = e, \mu, \tau$ corresponding to $j = 1, 2, 3$,
respectively.

From the global-fit data on neutrino mass-squared differences in
Eq.~(11) and the definition of $R^{}_\nu$ in Eq.~(12) one finds $
0.027 < R^{}_\nu < 0.037$ at the $3\sigma$ level. If $z^2_\nu \ll
x^2_\nu \ll 1$, then $R^{}_\nu \approx z^2_\nu /x^2_\nu$ holds as a
good approximation.

Due to the strong mass hierarchy of charged leptons the orthogonal
matrix $O^{}_l$ is approximately
\begin{equation}
O^{}_l \approx \left(\matrix{1 & - x^{1/2}_l & +z^{3/2}_l x^{-1}_l
\cr +x^{1/2}_l & 1 & +z^{1/2}_l x^{-1/2}_l \cr -z^{1/2}_l &
-z^{1/2}_l x^{-1/2}_l & 1}\right) \; .
%     (14)
\end{equation}
We obtain, using Eqs.~(8) and (14):
\begin{equation}
|V^{}_{e3}| < \sqrt{\frac{m^{}_1}{m^{}_2}} \cdot \left(\frac{\delta
m^2}{\Delta m^2}\right)^{3/4} + \sqrt{\frac{m^{}_e}{m^{}_\mu}} \cdot
\left(\frac{\delta m^2}{\Delta m^2}\right)^{1/4} +
\sqrt{\frac{m^{}_e}{m^{}_\tau}} \; ,
%     (15)
\end{equation}
where we have assumed $z^2_\nu \ll x^2_\nu \ll 1$, i.e. $R^{}_\nu
\approx z^2_\nu/x^2_\nu$. For $x^{}_\nu \approx 0.3$ and $R^{}_\nu <
0.037$ we then arrive at $|V^{}_{e3}| < 0.09$, which is in
contradiction with the lower bound $|V^{}_{e3}|
> 0.13$ at the $3\sigma$ level. It is clear from Eq.~(15) that
$\theta^{}_{13}$ is highly suppressed, because the ratio of two
neutrino mass-squared differences turns out to be small and the
contribution from the charged-lepton sector is negligible.

%%%%%%%%%%%%%%%%%%%%%%%%%%%%%%%%%%% Table 1 %%%%%%%%%%%%%%%%%%%%%%%%
\begin{table}[t]
\caption{The latest global-fit results of three neutrino mixing
angles $(\theta^{}_{12}, \theta^{}_{23}, \theta^{}_{13})$ and two
neutrino mass-squared differences $\delta m^2 \equiv m^2_2 - m^2_1$
and $\Delta m^2 \equiv m^2_3 - (m^2_1 + m^2_2)/2$ in the case of
normal neutrino mass hierarchy \cite{Fogli}.}
\begin{center}
\begin{tabular}{cccccc}
  \hline
  \hline
  Parameter & $\delta m^2~(10^{-5}~{\rm eV}^2)$ & $\Delta m^2~(10^{-3}~{\rm eV}^2)$
  & $\theta^{}_{12}$ & $\theta^{}_{23}$ & $\theta^{}_{13}$ \\
  \hline
  Best fit & $7.54$ & $2.43$ & $33.6^\circ$ & $38.4^\circ$ & $8.9^\circ$ \\
  $1\sigma$ range & $[7.32, 7.80]$ & $[2.33, 2.49]$ & $[32.6^\circ, 34.8^\circ]$
  & $[37.2^\circ, 40.0^\circ]$ & $[8.5^\circ, 9.4^\circ]$ \\
  $2\sigma$ range & $[7.15, 8.00]$ & $[2.27, 2.55]$ & $[31.6^\circ, 35.8^\circ]$
  & $[36.2^\circ, 42.0^\circ]$ & $[8.0^\circ, 9.8^\circ]$ \\
  $3\sigma$ range & $[6.99, 8.18]$ & $[2.19, 2.62]$ & $[30.6^\circ, 36.8^\circ]$
  & $[35.1^\circ, 53.0^\circ]$ & $[7.5^\circ, 10.2^\circ]$ \\
  \hline
\end{tabular}
\end{center}
\end{table}
%%%%%%%%%%%%%%%%%%%%%%%%%%%%%%%%%%%%%%%%%%%%%%%%%%%%%%%%%%%%%%%%%%%%%

Thus the scenario with both $M^{}_l$ and $M^{}_\nu$ of the form in
Eq.~(3) has been ruled out by current experimental data on neutrino
oscillations at the $3\sigma$ level.

\section{Canonical Seesaw Model}

In order to accommodate the tiny neutrino masses, one can extend the
standard model by introducing three right-handed singlet neutrinos.
The Lagrangian relevant for lepton masses is
\begin{equation}
-{\cal L}_{\rm SS} = \overline{l^{}_{\rm L}} Y^{}_l H E^{}_{\rm R} +
\overline{l^{}_{\rm L}} Y^{}_\nu \tilde{H} N^{}_{\rm R} +
\frac{1}{2} \overline{N^c_{\rm R}} M^{}_{\rm R} N^{}_{\rm R} + {\rm
h.c.} \; ,
%     (16)
\end{equation}
where $l^{}_{\rm L}$ and $\tilde{H} \equiv i\sigma^{}_2 H^*$ stand
respectively for the lepton and Higgs doublets, $E^{}_{\rm R}$ and
$N^{}_{\rm R}$ denote the charged-lepton and neutrino singlets,
$Y^{}_l$ and $Y^{}_\nu$ are the charged-lepton and Dirac neutrino
Yukawa coupling matrices, and $M^{}_{\rm R}$ is the mass matrix of
right-handed neutrinos.

After the electroweak gauge symmetry breaking the charged-lepton and
Dirac neutrino mass matrices are given by $M^{}_l = Y^{}_l v$ and
$M^{}_{\rm D} = Y^{}_\nu v$, with $v \approx 174~{\rm GeV}$ being
the vacuum expectation value of the Higgs field. The effective mass
matrix of the three light neutrinos is then determined by the seesaw
formula $M^{}_\nu = M^{}_{\rm D} M^{-1}_{\rm R} M^{\rm T}_{\rm D}$
\cite{type1}. Given ${\cal O}(M^{}_{\rm R}) \sim 10^{14}~{\rm GeV}$
close to the grand-unified-theory scale and ${\cal O}(M^{}_{\rm D})
\sim 10^2~{\rm GeV}$ at the electroweak scale, the neutrino masses
are in the sub-eV region. The smallness of neutrino masses can be
ascribed to the heaviness of right-handed Majorana neutrinos.
Although the seesaw mechanism can explain well the tiny neutrino
masses, it cannot fix the structure of lepton mass matrices that
determines the lepton masses and the flavor mixing pattern.

We assume that all the lepton mass matrices $M^{}_l$, $M^{}_{\rm D}$
and $M^{}_{\rm R}$ have the texture zeros, given in Eq.~(3), and
denote the corresponding matrix elements by different subscripts $f
= l, {\rm D}, {\rm R}$. Thus the effective neutrino mass matrix is
\begin{equation}
M^{}_\nu =  M^{}_{\rm D} M^{-1}_{\rm R} M^T_{\rm D} = \left(\matrix{
0 & {\cal C}^{}_\nu & 0 \cr {\cal C}^{}_\nu & {\cal D}^{}_\nu &
{\cal B}^{}_\nu \cr 0 & {\cal B}^{}_\nu & {\cal A}^{}_\nu}\right) \;
,
%     (17)
\end{equation}
where the matrix elements are given by
\begin{eqnarray}
{\cal A}^{}_\nu &=& \frac{{\cal A}^2_{\rm D}}{{\cal A}^{}_{\rm R}}
\; , \nonumber \\
{\cal C}^{}_\nu &=& \frac{{\cal C}^2_{\rm D}}{{\cal C}^{}_{\rm R}}
\; , \nonumber \\
{\cal D}^{}_\nu &=& \frac{{\cal C}^2_{\rm D}}{{\cal A}^{}_{\rm R}}
\left(\frac{{\cal B}^{}_{\rm D}}{{\cal C}^{}_{\rm D}} - \frac{{\cal
B}^{}_{\rm R}}{{\cal C}^{}_{\rm R}}\right)^2 \; , \nonumber \\
{\cal B}^{}_\nu &=& \frac{{\cal B}^{}_{\rm D} {\cal C}^{}_{\rm
D}}{{\cal C}^{}_{\rm R}} + \frac{{\cal C}^{}_{\rm D} {\cal
A}^{}_{\rm D}}{{\cal A}^{}_{\rm R}} \left(\frac{{\cal B}^{}_{\rm
D}}{{\cal C}^{}_{\rm D}} - \frac{{\cal B}^{}_{\rm R}}{{\cal
C}^{}_{\rm R}}\right) \; .
%     (18)
\end{eqnarray}
If the condition ${\cal B}^{}_{\rm D}/{\cal C}^{}_{\rm D} = {\cal
B}^{}_{\rm R}/{\cal C}^{}_{\rm R}$ is satisfied, then ${\cal
D}^{}_\nu = 0$ [see Eq.~(3)]. The matrix elements of $M^{}_\nu$
individually follow the seesaw relation ${\cal A}_\nu =  {\cal
A}^2_{\rm D}/{\cal A}^{}_{\rm R}$, ${\cal B}_\nu =  {\cal B}^2_{\rm
D}/{\cal B}^{}_{\rm R}$, and ${\cal C}_\nu = {\cal C}^2_{\rm
D}/{\cal C}^{}_{\rm R}$ \cite{XZ05}. Although this seesaw-invariant
scenario is phenomenologically interesting, it is disfavored by the
neutrino oscillation data, as we have shown in the previous section.
Another different model with texture zeros of the lepton mass
matrices in the seesaw framework can be found in Ref.~\cite{FTY}.

In order to obtain the leptonic mixing matrix, one must diagonalize
both the charged-lepton and effective neutrino mass matrices. The
charged-lepton mass matrix $M^{}_l$ can be diagonalized in the same
way as before, while the neutrino mass matrix $M^{}_\nu$ in Eq. (17)
can be diagonalized as follows. For simplicity we assume $\arg[{\cal
D}^{}_\nu] = 2\arg[{\cal B}^{}_\nu]$, then $M^{}_\nu$ can be
decomposed as $M^{}_\nu = P^{}_\nu \overline{M}^\prime_\nu P^T_\nu$
with
\begin{equation}
\overline{M}^\prime_\nu = \left(\matrix{ 0 & C^{}_\nu & 0 \cr
C^{}_\nu & D^{}_\nu & B^{}_\nu \cr 0 & B^{}_\nu & A^{}_\nu}\right)
\; ,
%     (19)
\end{equation}
and $P^{}_\nu = {\rm Diag}\{e^{i(\varphi^{}_\nu - \phi^{}_\nu)},
e^{i\phi^{}_\nu}, 1\}$, where $A^{}_\nu = {\cal A}^{}_\nu$,
$B^{}_\nu = |{\cal B}^{}_\nu|$, $C^{}_\nu = |{\cal C}^{}_\nu|$,
$D^{}_\nu = |{\cal D}^{}_\nu|$, $\varphi^{}_\nu = \arg[{\cal
C}^{}_\nu]$ and $\phi^{}_\nu = \arg[{\cal B}^{}_\nu]$. The real and
symmetric matrix $\overline{M}^\prime_\nu$ can be diagonalized by an
orthogonal transformation

\begin{equation}
O^T_\nu \overline{M}^\prime_\nu O^{}_\nu = \left(\matrix{
\lambda^\nu_1 & 0 & 0 \cr 0 & \lambda^\nu_2 & 0 \cr 0 & 0 &
\lambda^\nu_3}\right) \; ,
%     (20)
\end{equation}
where $\lambda^\nu_1 \lambda^\nu_2 < 0$ follows from ${\rm
Det}[\overline{M}^\prime_\nu] < 0$.

We shall discuss two different cases: $(\lambda^\nu_1,
\lambda^\nu_2) = (+m^{}_1, -m^{}_2)$ and $(\lambda^\nu_1,
\lambda^\nu_2) = (-m^{}_1, +m^{}_2)$. We take $(\lambda^\nu_1,
\lambda^\nu_2) = (+m^{}_1, -m^{}_2)$ and $\lambda^\nu_3 = m^{}_3 >
0$. The other case can be discussed in a similar way. The four
independent real parameters in $\overline{M}^\prime_\nu$ cannot be
expressed in terms of the three neutrino mass eigenvalues $m^{}_1$,
$m^{}_2$, and $m^{}_3$. Hence we define $r^{}_\nu \equiv
D^{}_\nu/A^{}_\nu$ and obtain
\begin{eqnarray}
A^{}_\nu &=& (m^{}_1 - m^{}_2 + m^{}_3)/(1+r^{}_\nu) \;, \nonumber
\\
B^{}_\nu &=& \left[\frac{(r^{}_\nu m^{}_1  + m^{}_2 - m^{}_3)(m^{}_1
+ r^{}_\nu m^{}_2  + m^{}_3)(m^{}_1 - m^{}_2 - r^{}_\nu m^{}_3
)}{(m^{}_1 - m^{}_2 + m^{}_3)(1+r^{}_\nu)^2}\right]^{1/2} \;
, \nonumber \\
C^{}_\nu &=& \left[\frac{m^{}_1 m^{}_2 m^{}_3 (1 + r^{}_\nu)}{m^{}_1
- m^{}_2 + m^{}_3}\right]^{1/2} \; .
%     (21)
\end{eqnarray}
The matrix elements of $O^{}_\nu$ are
\begin{eqnarray}
O^\nu_{11} &=& + \left[\frac{x^{}_\nu - z^{}_\nu - r^{}_\nu
x^2_\nu}{(1+x^{}_\nu)(1-z^{}_\nu)(x^{}_\nu - z^{}_\nu + x^{}_\nu
z^{}_\nu)}\right]^{1/2} \; , \nonumber \\
O^\nu_{12} &=& - \left[\frac{x^2_\nu ( x^{}_\nu + x^{}_\nu z^{}_\nu
+ r^{}_\nu z^{}_\nu )}{(1+x^{}_\nu)(x^{}_\nu + z^{}_\nu)(x^{}_\nu -
z^{}_\nu
+ x^{}_\nu z^{}_\nu)}\right]^{1/2} \; , \nonumber \\
O^\nu_{13} &=& + \left[\frac{z^2_\nu(z^{}_\nu - x^{}_\nu z^{}_\nu +
r^{}_\nu x^{}_\nu)}{(1-z^{}_\nu)(x^{}_\nu + z^{}_\nu)(x^{}_\nu -
z^{}_\nu + x^{}_\nu z^{}_\nu)}\right]^{1/2} \; , \nonumber \\
O^\nu_{21} &=& + \left[\frac{x^{}_\nu - z^{}_\nu + r^{}_\nu x^{}_\nu
z^{}_\nu}{(1+x^{}_\nu)(1-z^{}_\nu)(1 + r^{}_\nu)}\right]^{1/2} \; ,
\nonumber \\
O^\nu_{22} &=& + \left[\frac{x^{}_\nu + x^{}_\nu z^{}_\nu + r^{}_\nu
z^{}_\nu}{(1+x^{}_\nu)(x^{}_\nu +
z^{}_\nu)(1+r^{}_\nu)}\right]^{1/2} \; , \nonumber \\
O^\nu_{23} &=& + \left[\frac{z^{}_\nu - x^{}_\nu z^{}_\nu + r^{}_\nu
x^{}_\nu}{(1-z^{}_\nu)(x^{}_\nu +
z^{}_\nu)(1+r^{}_\nu)}\right]^{1/2} \; , \nonumber \\
O^\nu_{31} &=& - \left[\frac{(x^{}_\nu + x^{}_\nu z^{}_\nu +
r^{}_\nu z^{}_\nu)(z^{}_\nu - x^{}_\nu z^{}_\nu + r^{}_\nu
x^{}_\nu)}{(1+x^{}_\nu)(1-z^{}_\nu)(x^{}_\nu - z^{}_\nu + x^{}_\nu
z^{}_\nu)(1+r^{}_\nu)}\right]^{1/2} \; ,\nonumber \\
O^\nu_{32} &=& - \left[\frac{(x^{}_\nu - z^{}_\nu - r^{}_\nu
x^{}_\nu z^{}_\nu)(z^{}_\nu - x^{}_\nu z^{}_\nu + r^{}_\nu
x^{}_\nu)}{(1+x^{}_\nu)(x^{}_\nu + z^{}_\nu)(x^{}_\nu - z^{}_\nu +
x^{}_\nu z^{}_\nu)(1+r^{}_\nu)}\right]^{1/2} \; ,\nonumber \\
O^\nu_{33} &=& + \left[\frac{(x^{}_\nu - z^{}_\nu - r^{}_\nu
x^{}_\nu z^{}_\nu)(x^{}_\nu + x^{}_\nu z^{}_\nu + r^{}_\nu
z^{}_\nu)}{(1-z^{}_\nu)(x^{}_\nu + z^{}_\nu)(x^{}_\nu - z^{}_\nu +
x^{}_\nu z^{}_\nu)(1+r^{}_\nu)}\right]^{1/2} \; .
%     (22)
\end{eqnarray}

The lepton mixing matrix $V$ is given by Eq.~(8), however, the
matrix elements of $O^{}_\nu$ should be replaced by those in
Eq.~(22). The mixing matrix $V$ is entirely determined by the
charged-lepton mass ratios $(x^{}_l, z^{}_l)$, the neutrino mass
ratios $(x^{}_\nu, z^{}_\nu)$, the new parameter $r^{}_\nu$ and two
phases $(\alpha, \beta)$. With the additional parameter $r^{}_\nu$
we expect that the neutrino oscillation data can be explained. In
the following we consider two simplified textures of $M^{}_\nu$ and
illustrate, how the texture zeros of the lepton mass matrices
survive current experimental data:
\begin{itemize}
\item $D^{}_\nu = m^{}_2$, where $m^{}_2$ denotes the mass of
$\nu^{}_2$. This assumption has conventionally been made in the
study of the four-zero textures of quark \cite{Du} and lepton mass
matrices \cite{Zhang}.

\item $D^{}_\nu = A^{}_\nu$. It is worthwhile to note that the $2$-$3$
block of neutrino mass matrix $M^{}_\nu$ in this case can be
diagonalized by a maximal rotation angle. However, the mixing angle
$\theta^{}_{23}$ should not be maximal because of the moderate
neutrino mass hierarchy and the correction from the charged-lepton
sector.
\end{itemize}

As mentioned before, $D^{}_\nu = 0$ if ${\cal B}^{}_{\rm D}/{\cal
C}^{}_{\rm D} = {\cal B}^{}_{\rm R}/{\cal C}^{}_{\rm R}$ holds, so
both the charged-lepton and effective neutrino mass matrices in this
case are of the form, given in Eq.~(3). This scenario is not
consistent with the neutrino oscillation data, which indicates that
the structure of Dirac neutrino mass matrix $M^{}_{\rm D}$ should be
quite different from that of the heavy Majorana neutrino mass matrix
$M^{}_{\rm R}$. This seems to be more natural because the former is
generally an arbitrary matrix, while the latter should be symmetric.

\subsection{Case (A): $D^{}_\nu = m^{}_2$}

In this case we obtain $r^{}_\nu = z^{}_\nu/(x^{}_\nu - 2 z^{}_\nu +
x^{}_\nu z^{}_\nu)$. Substituting $r^{}_\nu$ into Eq.~(21), one can
calculate the other non-vanishing matrix elements of
$\overline{M}^\prime_\nu$ in Eq.~(19)
\begin{eqnarray}
A^{}_\nu &=& m^{}_3 - 2m^{}_2 + m^{}_1 \; , \nonumber \\
B^{}_\nu &=& \left[\frac{(m^{}_3 - 2m^{}_2)(m^{}_3 - m^{}_2 + m^{}_1
)(2m^{}_2 - m^{}_1)}{m^{}_3 - 2m^{}_2 + m^{}_1}\right]^{1/2} \; ,
\nonumber \\
C^{}_\nu &=& \left[\frac{m^{}_1 m^{}_2 m^{}_3}{m^{}_3 - 2m^{}_2 +
m^{}_1}\right]^{1/2} \; .
%     (23)
\end{eqnarray}

The matrix elements of $\overline{M}^{}_l$ in Eq.~(4) are given by
\begin{eqnarray}
A^{}_l &=& m^{}_\tau - m^{}_\mu + m^{}_e \; , \nonumber \\
B^{}_l &=& \left[\frac{(m^{}_\mu - m^{}_e)(m^{}_\tau -
m^{}_\mu)(m^{}_\tau + m^{}_e)}{m^{}_\tau - m^{}_\mu +
m^{}_e}\right]^{1/2} \; , \nonumber \\
C^{}_l &=& \left[\frac{m^{}_e m^{}_\mu m^{}_\tau}{m^{}_\tau -
m^{}_\mu + m^{}_e}\right]^{1/2} \; .
%     (24)
\end{eqnarray}
Approximately one finds $A^{}_l \approx m^{}_\tau$, $B^{}_l \approx
\sqrt{m^{}_\mu m^{}_\tau}$, and $C^{}_l \approx \sqrt{m^{}_e
m^{}_\mu}$. Inserting $r^{}_\nu$ into Eq.~(22), we can obtain the
orthogonal matrix $O^{}_\nu$ and the lepton mixing matrix $V$, which
consists of four parameters $(x^{}_\nu, z^{}_\nu)$ and $(\alpha,
\beta)$. By using the current neutrino oscillation data, we find
that {\bf Case (A)} is successful at the $3\sigma$ level. The
allowed regions of the neutrino mass ratios $(x^{}_\nu, z^{}_\nu)$
and the phases $(\alpha, \beta)$, together with those of the
neutrino mixing parameters and other observables, are shown in
Fig.~1. We note:

\begin{enumerate}
\item The allowed regions of $(x^{}_\nu, z^{}_\nu)$ and $(\alpha, \beta)$
are given in the two plots in the first row of Fig.~1. We observe
that $x^{}_\nu \sim 0.5$ and $z^{}_\nu \sim 0.1$ are typical values,
so the neutrinos have a normal mass hierarchy $m^{}_1 < m^{}_2 <
m^{}_3$. Note that both the neutrino mass ratios $(x^{}_\nu,
z^{}_\nu)$ and the phase parameters $(\alpha, \beta)$ are
restrictively constrained. Only a fraction of the parameter space
around $\alpha = \pi$ and $\beta = \pi$ is allowed.

\item In the second row of Fig.~1 we show the predictions for the
neutrino mixing angles $(\theta^{}_{12}, \theta^{}_{23},
\theta^{}_{13})$ and the ratio of neutrino mass-squared differences
$R^{}_\nu$. Although a relatively large $\theta^{}_{13}$ can now be
achieved, an upper bound $\theta^{}_{13} < 8.8^\circ$ exists, an
even larger $\theta^{}_{13}$ requires a large value of $R^{}_\nu$
that is experimentally disfavored. Moreover $\theta^{}_{12} >
32.6^\circ$ and $\theta^{}_{23} < 46.5^\circ$ are predicted. In the
same plots the best-fit values of $(\theta^{}_{12}, \theta^{}_{23})$
and $(R^{}_\nu, \theta^{}_{13})$ are represented by red solid
squares, which are lying far away from the allowed parameter space.
The $3\sigma$, $2\sigma$ and $1\sigma$ ranges are bounded by the
solid, dashed and dotted lines respectively. Therefore more precise
measurements of neutrino mixing parameters are needed to verify or
rule out the model under consideration.

\item The predictions for the Jarlskog invariant $J$ and the
Dirac-type and Majorana-type CP-violating phases $(\delta, \rho,
\sigma)$ are given in the third row of Fig.~1. If the CP-violating
phase takes the maximal value of $\delta = 0.18~\pi$, then the
Jarlskog invariant $J$ can reach $1.6\%$, which could hopefully be
measured in future long-baseline neutrino oscillation experiments.
On the other hand, two Majorana-type CP-violating phases are found
to be $\rho \sim 0$ and $\sigma \sim \pi/2$. The relation $\sigma -
\rho = \pi/2$ holds due to the imaginary unit arising from ${\rm
Det} [M^{}_\nu] < 0$, and $\alpha \sim \beta \sim \pi$, as shown in
Fig.~1 .

\item The last row of Fig.~1 shows the predictions for the absolute
neutrino mass $m^{}_3$, the effective neutrino mass $\langle m
\rangle^{}_\beta$ in the tritium beta decay, and the effective
neutrino mass $\langle m \rangle^{}_{\beta \beta}$ in the
neutrinoless double-beta decays. One can observe that $\langle  m
\rangle^{}_\beta < 1.0\times 10^{-2}~{\rm eV}$ and $\langle m
\rangle^{}_{\beta \beta} < 2.4\times 10^{-3}~{\rm eV}$, both of
which are difficult to be measured in the ongoing experiments. This
is the usual situation for the case of normal neutrino mass
hierarchy, in which the contribution from the heaviest neutrino mass
eigenstate $\nu^{}_3$ is suppressed by the smallest mixing angle
$\theta^{}_{13}$.
\end{enumerate}

We conclude that {\bf Case (A)} is compatible with the current
neutrino oscillation data at the $3\sigma$ level. The future
precision measurements of the neutrino mixing parameters can rule
out this case if the experimental results finally converge to the
current best-fit values.

Taking the typical values of $m^{}_3 = 5.0\times 10^{-2}~{\rm eV}$
and $(x^{}_\nu, z^{}_\nu) = (0.5, 0.1)$, one can calculate the other
neutrino masses: $m^{}_2 = 1.0\times 10^{-2}~{\rm eV}$ and $m^{}_1 =
5.0\times 10^{-3}~{\rm eV}$. One can also compute the matrix
elements of $\overline{M}^\prime_\nu$ and $\overline{M}^{}_l$:
\begin{eqnarray}
\overline{M}^\prime_\nu & \approx & 3.5\times 10^{-2}~{\rm eV} \cdot
\left(\matrix{0 & 0.24& 0 \cr 0.24 & 0.29 & 0.69 \cr 0 & 0.69 & 1}\right)\; , \nonumber \\
\overline{M}^{}_l &\approx& 1.67~{\rm GeV} \cdot \left(\matrix{0 &
0.0045 & 0 \cr 0.0045 & 0 & 0.26 \cr 0 & 0.26 & 1}\right)\; .
\end{eqnarray}
In order to see the structure of the heavy Majorana neutrino mass
matrix $M^{}_{\rm R}$, we have to specify the structure of Dirac
neutrino mass matrix $M^{}_{\rm D}$. Quite different from the
four-zero textures of lepton mass matrices in the seesaw model, the
Dirac neutrino mass matrix $M^{}_{\rm D}$ is not arbitrary, but
subject to the constraint relation
\begin{equation}
\frac{B^{}_{\rm D}}{C^{}_{\rm D}} = \frac{B^{}_\nu}{C^{}_\nu} +
\frac{A^{}_\nu}{C^{}_\nu} \sqrt{\frac{D^{}_\nu}{A^{}_\nu}}
\end{equation}
from Eq.~(18), where we neglect the phases of the matrix elements
for an order of magnitude estimate. Since $M^{}_{\rm D} $ is of the
form in Eq.~(3), we write it in terms of its eigenvalues:
\begin{equation}
M^{}_{\rm D} = \left(\matrix{0 & \sqrt{d^{}_1 d^{}_2} & 0 \cr
\sqrt{d^{}_1 d^{}_2} & 0 & \sqrt{d^{}_2 d^{}_3} \cr 0 & \sqrt{d^{}_2
d^{}_3} & d^{}_3}\right)
\end{equation}
where $d^{}_3 = m^{}_t = 174~{\rm GeV}$ is the running top-quark
mass at the electroweak scale. For illustration $d^{}_1 : d^{}_2 :
d^{}_3 = 1 : 5 : 25$ will be assumed to satisfy the constraint in
Eq.~(26). Thus $M^{}_{\rm R}$ is found via the seesaw formula
$M^{}_{\rm R} = M^T_{\rm D} M^{-1}_\nu M^{}_{\rm D}$ to be
\begin{equation}
M^{}_{\rm R} \approx 8.7\times 10^{14}~{\rm GeV} \cdot
\left(\matrix{0 & 0.033 & 0 \cr 0.033 & 0 & 0.35 \cr 0 & 0.35 &
1}\right) \; .
\end{equation}

It is obvious that both $M^{}_{\rm D}$ and $M^{}_{\rm R}$ cannot be
strongly hierarchical, although the hierarchy of $M^{}_{\rm R}$ is
slightly stronger than that of $M^{}_{\rm D}$. With the help of
Eq.~(18) we find that the hierarchy of $M^{}_{\rm R}$ is dictated by
$M^{}_{\rm D}$ and $M^{}_\nu$, since
\begin{equation}
\frac{B^{}_{\rm R}}{C^{}_{\rm R}} = \frac{B^{}_{\rm D}}{C^{}_{\rm
D}} + \frac{A^{}_{\rm D}}{C^{}_{\rm D}}
\sqrt{\frac{D^{}_\nu}{A^{}_\nu}} \; .
\end{equation}
Note that the relation $B^{}_\nu/C^{}_\nu = B^{}_{\rm D}/C^{}_{\rm
D} = B^{}_{\rm R}/C^{}_{\rm R}$ is reproduced if we set $D^{}_\nu =
0$ in Eqs.~(26) and (29). If the complex phases in $M^{}_{\rm R}$
and $M^{}_{\rm D}$ are included, the lepton number asymmetry can be
produced in the CP-violating and out-of-equilibrium decays of heavy
Majorana neutrinos in the early Universe \cite{FY}. Through the
sphaleron processes, the lepton number asymmetry will be converted
into baryon number asymmetry, explaining the matter-antimatter
asymmetry in our Universe. We expect this elegant leptogenesis
mechanism works well in the present case.

\subsection{Case (B): $D^{}_\nu = A^{}_\nu$}

Now we consider another example with $D^{}_\nu = A^{}_\nu$, namely
$r^{}_\nu = 1$. In this case the other non-vanishing matrix elements
of $\overline{M}^\prime_\nu$ in Eq.~(19) are:
\begin{eqnarray}
A^{}_\nu &=& \frac{1}{2} (m^{}_1 - m^{}_2 + m^{}_3) \;, \nonumber
\\
B^{}_\nu &=& \frac{1}{2}\left[\frac{(m^{}_1 + m^{}_2 -
m^{}_3)(m^{}_1 + m^{}_2 + m^{}_3)(m^{}_1 - m^{}_2 - m^{}_3
)}{(m^{}_1 - m^{}_2 + m^{}_3)}\right]^{1/2} \;
, \nonumber \\
C^{}_\nu &=& \left[\frac{2m^{}_1 m^{}_2 m^{}_3}{m^{}_1 - m^{}_2 +
m^{}_3}\right]^{1/2} \; .
%     (30)
\end{eqnarray}
The matrix elements of $\overline{M}^{}_l$ are still given by
Eq.~(24). Inserting $r^{}_\nu = 1$ into Eq.~(22), we obtain the
orthogonal matrix $O^{}_\nu$ and the lepton mixing matrix $V$, which
is completely determined by four parameters $(x^{}_\nu, z^{}_\nu)$
and $(\alpha, \beta)$. With the help of current neutrino oscillation
data, we have found ${\bf Case~(B)}$ is perfectly consistent with
experimental data -- even the best-fit values of neutrino mixing
parameters can be reproduced. Our numerical results are shown in
Fig.~2. We note:

\begin{enumerate}
\item The allowed parameter space of the neutrino mass ratios
$(x^{}_\nu, z^{}_\nu)$ and the phase parameters $(\alpha, \beta)$ is
given in the plots in the first row of Fig. 2. The typical values of
the neutrino mass ratios are $x^{}_\nu = 0.32$ and $z^{}_\nu =
0.06$, indicating a moderate mass hierarchy $m^{}_1 : m^{}_2 :
m^{}_3 \approx 1 : 3 : 15$. Note that $\beta = \pi$ is
experimentally disallowed, and only a fraction of parameter space
around $\alpha \sim \beta \sim \pi/2$ (or $3\pi/2$) is favored.

\item The predictions for three neutrino mixing angles $(\theta^{}_{12},
\theta^{}_{23}, \theta^{}_{13})$ and the ratio of the neutrino
mass-squared differences $R^{}_\nu$ are shown in the second row of
Fig.~2. One can observe that even the best-fit values (red squares)
and the $1\sigma$ ranges (regions between the dotted lines) of these
observables can be achieved. To understand this result, we calculate
$|V^{}_{e3}|$ in the same way as in Eq.~(15):
\begin{equation}
|V^{}_{e3}| < \sqrt{\frac{m^{}_1}{m^{}_2}} \left(\frac{\delta
m^2}{\Delta m^2}\right)^{1/2} +
\frac{1}{\sqrt{2}}\left(\sqrt{\frac{m^{}_e}{m^{}_\mu}} +
\sqrt{\frac{m^{}_e}{m^{}_\tau}}\right)\; ,
\end{equation}
where we have assumed $z^2_\nu \ll x^2_\nu \ll 1$, i.e. $R^{}_\nu
\approx z^2_\nu/x^2_\nu$. Given $x^{}_\nu \approx 0.3$ and $R^{}_\nu
< 0.037$ at the $3\sigma$ level, we arrive at $|V^{}_{e3}| < 0.167$,
which is well compatible with the current neutrino oscillation data.

\item As shown in the plots in the last two rows of Fig.~2, the
predictions for the leptonic CP violation, the effective neutrino
masses in the tritium decay and the neutrinoless double-beta decay
are similar to those in ${\bf Case~(A)}$. It is possible to discover
leptonic CP violation in the future long-baseline neutrino
oscillation experiments, while it is quite challenging to determine
the Majorana neutrino mass in the neutrinoless double-beta decay.
\end{enumerate}

Taking the typical values of $m^{}_3 = 5.0\times 10^{-2}~{\rm eV}$
and $(x^{}_\nu, z^{}_\nu) = (0.32, 0.06)$, one can determine the
other neutrino masses: $m^{}_2 = 9.4\times 10^{-3}~{\rm eV}$ and
$m^{}_1 = 3.0\times 10^{-3}~{\rm eV}$. For an order of magnitude
estimate we neglect the complex phases in the lepton mass matrices.
The neutrino and charged-lepton mass matrices are:
\begin{eqnarray}
\overline{M}^\prime_\nu & \approx & 2.2\times 10^{-2}~{\rm eV} \cdot
\left(\matrix{0 & 0.37& 0 \cr 0.37 & 1 & 1.26 \cr 0 & 1.26 &
1}\right)
\; , \nonumber \\
\overline{M}^{}_l &\approx& 1.67~{\rm GeV} \cdot \left(\matrix{0 &
0.0045 & 0 \cr 0.0045 & 0 & 0.26 \cr 0 & 0.26 & 1}\right)\; .
\end{eqnarray}
Note that $B^{}_\nu > A^{}_\nu$ holds for the effective neutrino
mass matrix. In order to see the structure of $M^{}_{\rm R}$, we
assume $M^{}_{\rm D}$ to be the same as in Eq.~(27), but with
$d^{}_1 : d^{}_2 : d^{}_3 = 1 : 6 : 36$, and obtain
\begin{equation}
M^{}_{\rm R} \approx 1.4\times 10^{15}~{\rm GeV} \cdot
\left(\matrix{0 & 0.01 & 0 \cr 0.01 & 0 & 0.25 \cr 0 & 0.25 &
1}\right) \; .
\end{equation}
Again both $M^{}_{\rm D}$ and $M^{}_{\rm R}$ cannot have a strong
hierarchy in the matrix elements, due to the moderate neutrino mass
hierarchy and the constraint conditions in Eqs.~(26) and (29).

\section{Summary}

Taking into account the new results of the neutrino oscillation
experiments, we study the validity of the texture zeros in the
lepton mass matrices. Due to the large angle $\theta^{}_{13}$ the
original scenario, in which both the charged-lepton mass matrix
$M^{}_l$ and the effective neutrino mass matrix $M^{}_\nu$ have
three texture zeros, is inconsistent with current neutrino
oscillation data. In the canonical seesaw model we apply the texture
zeros to the charged-lepton mass matrix $M^{}_l$, the Dirac neutrino
mass matrix $M^{}_{\rm D}$, and the heavy Majorana neutrino mass
matrix $M^{}_{\rm R}$. We present two phenomenologically interesting
models that are compatible with the experimental data. The
nonmaximal $\theta^{}_{23}$ and unsuppressed $\theta^{}_{13}$ can
naturally be accommodated. The neutrinos have a normal mass
hierarchy, and the effective neutrino masses $\langle m
\rangle^{}_\beta$ in the tritium beta decay and $\langle m
\rangle^{}_{\beta \beta}$ in the neutrinoless double-beta decays are
found to be at the meV level. The future precision measurements of
the neutrino mixing parameters and the possible discovery of
leptonic CP violation will allow us to distinguish between these two
models and to test their validity.

\vspace{0.5cm}
\begin{flushleft}
{\large \bf Acknowledgements}
\end{flushleft}

One of us (S.Z.) is indebted to Zhi-zhong Xing for stimulating
discussions and helpful suggestions. This work was supported by the
G\"{o}ran Gustafsson Foundation.

\newpage

%%%%%%%%%%%%%%%%%%%% Fig. 1 %%%%%%%%%%%%%%%%
\begin{figure}
\epsfig{file=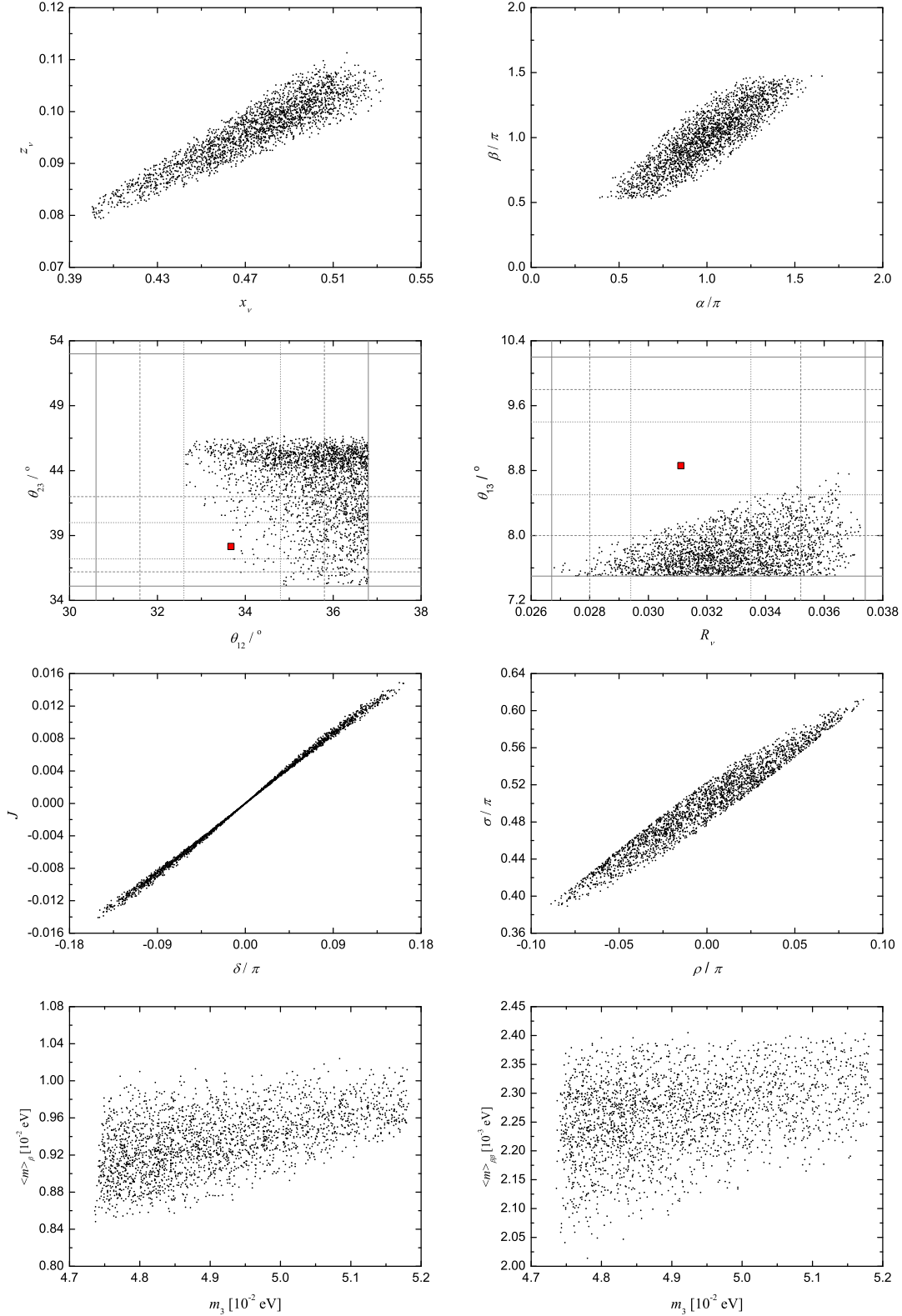,bbllx=2cm,bblly=11.5cm,bburx=17cm,bbury=26.5cm,%
width=15cm,height=15cm,angle=0,clip=0} \vspace{5.8cm}\caption{{\bf
Case (A)} with $D^{}_\nu = m^{}_2$: Allowed regions of the neutrino
mass ratios $(x^{}_\nu, z^{}_\nu)$ and two phase parameters
$(\alpha, \beta)$, where the $3\sigma$ ranges of neutrino mixing
angles and mass-squared differences are taken as input \cite{Fogli}.
The predictions for three neutrino mixing angles $(\theta^{}_{12},
\theta^{}_{23}, \theta^{}_{13})$, the CP-violating phases $(\delta,
\rho, \sigma)$, the Jarlskog invariant $J$, the effective neutrino
masses in the tritium beta decay $\langle m \rangle^{}_\beta$ and in
the neutrinoless double-beta decays $\langle m \rangle^{}_{\beta
\beta}$ are also shown.}
\end{figure}
%%%%%%%%%%%%%%%%%%%%%%%%%%%%%%%%%%%%%%%%%%%

%%%%%%%%%%%%%%%%%%%% Fig. 2 %%%%%%%%%%%%%%%%
\begin{figure}
\epsfig{file=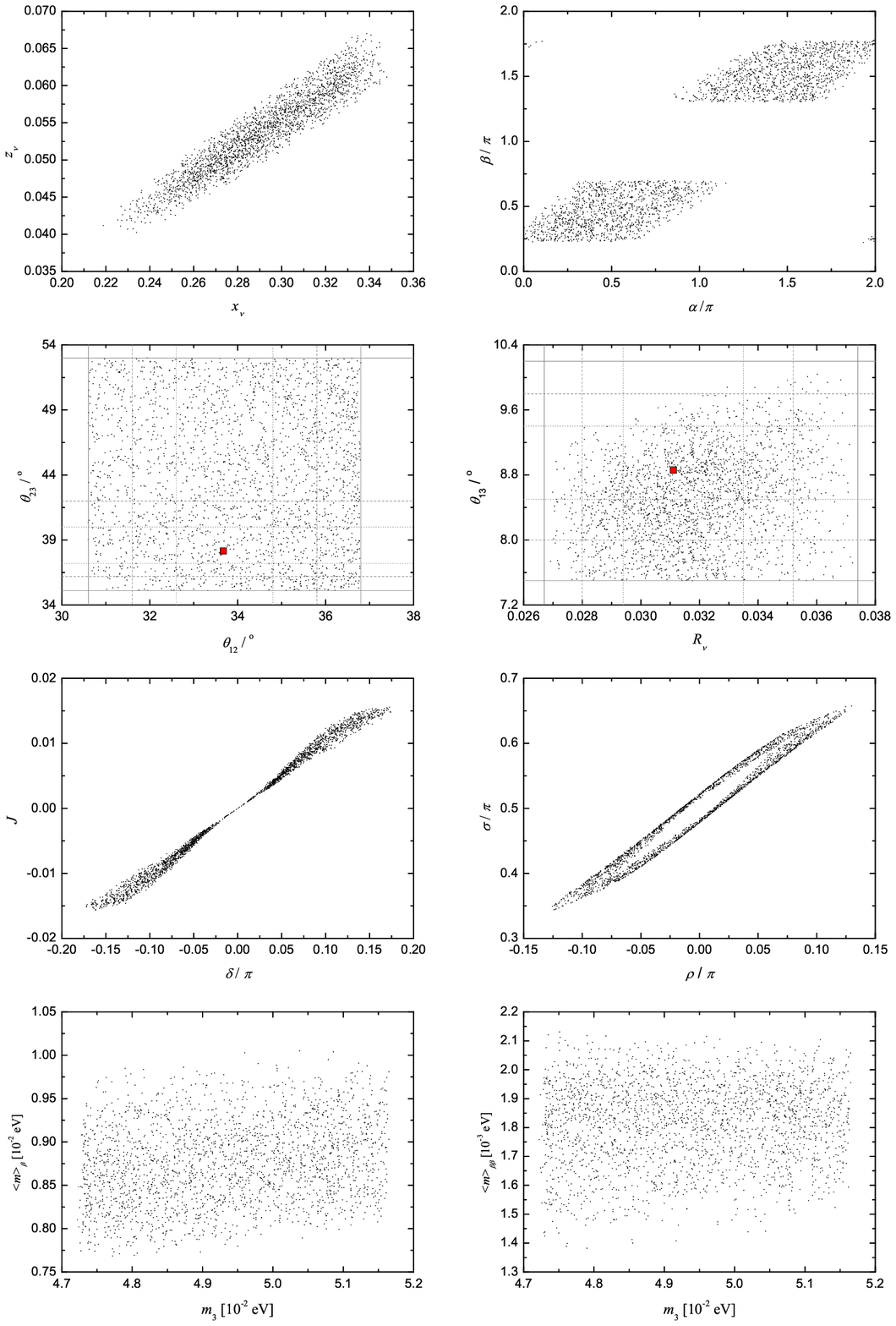,bbllx=2cm,bblly=11.5cm,bburx=17cm,bbury=26.5cm,%
width=15cm,height=15cm,angle=0,clip=0} \vspace{5.8cm}\caption{{\bf
Case (B)} with $D^{}_\nu = A^{}_\nu$: Allowed regions of the
neutrino mass ratios $(x^{}_\nu, z^{}_\nu)$ and two phase parameters
$(\alpha, \beta)$, where the $3\sigma$ ranges of neutrino mixing
angles and mass-squared differences are taken as input \cite{Fogli}.
The predictions for three neutrino mixing angles $(\theta^{}_{12},
\theta^{}_{23}, \theta^{}_{13})$, the CP-violating phases $(\delta,
\rho, \sigma)$, the Jarlskog invariant $J$, the effective neutrino
masses in the tritium beta decay $\langle m \rangle^{}_\beta$ and in
the neutrinoless double-beta decays $\langle m \rangle^{}_{\beta
\beta}$ are also shown.}
\end{figure}
%%%%%%%%%%%%%%%%%%%%%%%%%%%%%%%%%%%%%%%%%%%


\vspace{0.5cm}

\begin{thebibliography}{99}

\bibitem{Daya} F.P. An {\it et al.} (Daya Bay Collaboration), Phys.
Rev. Lett. {\bf 108}, 171803 (2012).

\bibitem{Reno} J.K. Ahn {\it et al.} (RENO Collaboration),
Phys. Rev. Lett. {\bf 108}, 191802 (2012).

\bibitem{F} H. Fritzsch, Phys. Lett. B {\bf 73}, 317 (1978);
Nucl. Phys. B {\bf 155}, 189 (1979).

\bibitem{Xing02} Z.Z. Xing, Phys. Lett. B {\bf 550}, 178 (2002).

\bibitem{XZ04} Z.Z. Xing and S. Zhou, Phys. Lett. B {\bf 593}, 156
(2004); S. Zhou and Z.Z. Xing, Eur. Phys. J. C {\bf 38}, 495 (2005).

\bibitem{Fogli} G.L. Fogli, E. Lisi, A. Marrone, D. Montanino, A.
Palazzo, and A.M. Rotunno, Phys. Rev. D {\bf 86}, 013012 (2012).

\bibitem{type1} H. Fritzsch, M. Gell-Mann, and P. Minkowski, Phys.
Lett. B {\bf 59}, 256 (1975); P. Minkowski, Phys. Lett. B {\bf 67},
421 (1977); T. Yanagida, in {\it Proceedings of the Workshop on
Unified Theory and the Baryon Number of the Universe}, edited by O.
Sawada and A. Sugamoto (KEK, Tsukuba, 1979), p. 95; M. Gell-Mann, P.
Ramond, and R. Slansky, in {\it Supergravity}, edited by P. van
Nieuwenhuizen and D.Z. Freeman (North-Holland, Amsterdam, 1979), p.
315; S.L. Glashow, in {\it Quarks and Leptons}, edited by M. Levy
{\it et al.} (Plenum, New York, 1980), p. 707; R.N. Mohapatra and G.
Senjanovic, Phys. Rev. Lett. {\bf 44}, 912 (1980).

\bibitem{XZ05} Z.Z. Xing and S. Zhou, Phys. Lett. B {\bf 606}, 145
(2005).

\bibitem{FTY} M. Fukugita, M. Tanimoto, and T. Yanagida,
Phys. Lett. B {\bf 562}, 273 (2003); M. Fukugita, Y. Shimizu, M.
Tanimoto, and T.T. Yanagida, Phys. Lett. B {\bf 716}, 294 (2012).

\bibitem{Du} D. Du and Z.Z. Xing, Phys. Rev. D {\bf 48}, 2349 (1993);
L.J. Hall and A. Rasin, Phys. Lett. B {\bf 315}, 164 (1993); H.
Fritzsch and D. Holtmansp\"{o}ter, Phys. Lett. B {\bf 338}, 290
(1994); H. Fritzsch and Z.Z. Xing, Phys. Lett. B {\bf 555}, 63
(2003); Z.Z. Xing and H. Zhang, J. Phys. G {\bf 30}, 129 (2004).

\bibitem{Zhang} Z.Z. Xing and H. Zhang, Phys. Lett. B {\bf 569}, 30
(2003).

\bibitem{FY} M. Fukugita and T. Yanagida, Phys. Lett. B {\bf 174}, 45
(1986).

\end{thebibliography}
\end{document}